\documentclass[useAMS,usenatbib]{mnras}
\usepackage{graphicx}
\usepackage{amsmath}
\usepackage{amssymb}
\usepackage{color}

\def \etal {et~al.~}
\def \msun {\ifmmode \rm M_{\odot} \else $\rm M_{\odot}$ \fi}
\newcommand{\Mpch}{{\ifmmode{h^{-1}{\rm Mpc}}\else{$h^{-1}$Mpc}\fi}}
\newcommand{\kms}{{\ifmmode{ {\rm km\,s^{-1}} }\else{ ${\rm km\,s^{-1}}$ }\fi}}
\def \hi {{\sc Hi} }
\newcommand{\Vmax}{\ifmmode{V_{\rm max}^{\rm DM}}\else{$V_{\rm max}^{\rm DM}$}\fi}

\title[The NIHAO velocity function]{NIHAO X: Reconciling the local galaxy velocity function  with Cold Dark Matter via mock \hi observations}
\author[Macci\`o \etal]{Andrea V. Macci\`{o}$^{1,2}$\thanks{E-mail: maccio@nyu.edu}, Silviu M. Udrescu$^1$,  Aaron A. Dutton$^{1}$, Aura Obreja$^{1}$, 
  \newauthor Liang Wang$^3$,  Greg R. Stinson$^2$, Xi Kang$^3$\\  
 $^{1}$New York University Abu Dhabi, PO Box 129188, Saadiyat Island, Abu Dhabi, United Arab Emirates\\
 $^{2}$Max Planck Institute f\"{u}r Astronomie, K\"{o}nigstuhl 17, 69117 Heidelberg, Germany\\
 $^{3}$Purple Mountain Observatory, the Partner Group of MPI f\"{u}r Astronomie, 2 West Beijing Road, Nanjing 210008, China\\
}
\begin{document}

\maketitle

\label{firstpage}

\begin{abstract}
  We used 87 high resolution hydrodynamical cosmological simulations
  from the NIHAO suite to investigate the relation between the maximum
  circular velocity (\Vmax) of a dark matter halo in a collisionless
  simulation and the velocity width of the \hi gas in the same halo in
  the hydrodynamical simulation. These two quantities are normally
  used to compare theoretical and observational velocity functions and
  have led to a possible discrepancy between observations and
  predictions based on the Cold Dark Matter (CDM) model. We show that
  below 100 $\kms$, there is clear bias between \hi based velocities
  and \Vmax, that leads to an underestimation of the actual circular
  velocity of the halo. When this bias is taken into account the CDM
  model has no trouble in reproducing the observed velocity function
  and no lack of low velocity galaxies is actually present. Our
  simulations also reproduce the linewidth - stellar mass
  (Tully-Fisher) relation and \hi sizes, indicating that the \hi gas
  in our simulations is as extended as observed.  The physical reason
  for the lower than expected linewidths is that, in contrast to high
  mass galaxies, low mass galaxies no longer have extended thin \hi rotating disks,
  as is commonly assumed. 
\end{abstract}


\vspace{-0.4cm}
\section{Introduction}

The Cold Dark Matter (CDM) scenario is very successful in reproducing
the large scale structure of the Universe \citep{Springel2005}
and the anisotropies in the Cosmic Microwave Background \citep{Planck2014}.
In a CDM universe structure formation
proceeds in a bottom-up fashion, with small structures collapsing
earlier  and then merging to form larger and larger haloes.
This scenario predicts a larger number of low mass structures compared
to more massive ones,  or in other words a steeply declining halo mass
function.  Such a prediction seems to be at odds with observational
data both around galaxies, e.g. the satellite abundance
\citep{Klypin1999, Moore1999} and in the local volume \citep{Cole1989,
  Zavala2009, Trujillo-Gomez2011}.

In the last years several solutions have been proposed to reconcile
the observed number of satellites around the Milky Way and
Andromeda galaxies with predictions from CDM \citep[][and
  references therein]{Bullock2000, Benson2002,  Maccio2010, Font2011}.  All these solutions are based (to a
different extent) on the fact that galaxy formation becomes quite
inefficient in low mass dark matter haloes, due to ionizing
background, Super Novae (SN) explosions and gas removal due to ram pressure. As a
result, by taking into account baryonic processes it is possible to
bring into agreement the abundance of satellites in the Local Group
and CDM predictions \citep[e.g.,][]{Sawala2016}.

A more persistent challenge is provided by field galaxies in the local
volume and their velocity function.  The galaxy velocity function
(GVF) is defined as the abundance of galaxies with a given circular
velocity and is conceptually similar to the galaxy luminosity
function. The advantage is that, for a given cosmological model,
circular velocities can nominally be predicted much more accurately than
luminosities, since they are less dependent on the still poorly
understood physics of galaxy formation. This makes velocity functions a very useful
tool to test theoretical models \citep{Zavala2009, Papastergis2011}.

Recently \citet[][hereafter K15]{Klypin2015} have presented new
measurements of the GVF in the Local Volume ($ D \approx 10$ Mpc) and
they have shown that while CDM provides very good
estimates of the number of galaxies with circular velocities around
and above $70\,\kms$, it fails quite dramatically at lower circular
velocities, overestimating by a factor up to five the number of dwarf
galaxies in the velocity range $30-50\,\kms$.
As pointed out by K15 and other authors \citep[e.g.,][]{Zavala2009}
galaxies in this velocity range are practically insensitive to
the ionization background and, by not being satellites, they are not
affected by gas depletion via ram pressure or by stellar stripping. This
makes the mismatch between the observed GVF and
the CDM predictions a quite serious problem for the current
cosmological model.  Moreover, simple modifications to the CDM
paradigm, for example by introducing a warm dark matter component
\citep[e.g.,][]{Schneider2014}, seem also to be not able to fix this
issue, as shown by K15.

\citet{Brook2015} showed that a population of galaxies with mass
profiles modified by baryonic feedback \citep{DiCintio2014b} is able
to explain the GVF significantly better than a model in which a
universal cuspy density profile is assumed.
In this Letter we revise
the effect of galaxy formation on the GVF using the hydrodynamical
simulations from the NIHAO project \citep{Wang2015}. In contrast to
many previous theoretical studies, which use proxies for \hi
linewidths, we directly measure them.  The combination of high spatial
resolution and large sample size make NIHAO ideally suited to study
the relation between the (rotation) velocity inferred by the
kinematics of the \hi gas component and the maximum circular velocity
of the halo, and thus to shed light on the reasons behind the large
discrepancy between the CDM-based predicted and the observed GVF.

\vspace{-0.5cm}
\section{Simulations}
\label{sec:sims}

The NIHAO project is a suite of fully cosmological hydrodynamical
``zoom-in'' simulations that covers a large range of galaxy masses from dwarf
galaxies to massive spirals (halo virial velocity from $25$ to $200\,\kms$).
 One characteristic of NIHAO galaxies is that they have roughly the
 same relative resolution and hence the same number of dark matter
 particles within the virial radius \cite[See Fig.~2 of][]{Wang2015}.
 The zoomed initial conditions were created using a modified version
 of {\sc grafic2} \citep{Bertschinger2001, Penzo2014}.
Each halo is initially simulated at high resolution with DM-only,
using {\sc pkdgrav} \citep{Stadel2001} then re-simulated with baryons.
We refer to the DM-only simulations as  {\it N-body} and the
simulations with baryons as {\it hydro} or NIHAO.

The hydrodynamical simulations are evolved using an improved version
of the SPH code {\sc gasoline} \citep{Wadsley2004, Keller2014}.  The code
includes a subgrid model for turbulent mixing of metals and energy
, heating and cooling include photo-electric
heating of dust grains, ultraviolet (UV) heating and ionization and
cooling due to hydrogen, helium and metals \citep{Shen2010}.
The star formation and feedback modeling follows what was used in the
MaGICC simulations \citep{Stinson2013}, adopting a threshold for star
formation of $n_{\rm th} > 10.3$ cm$^{-3}$.  Stars can feed energy
back into the ISM via blast-wave supernova (SN) feedback
\citep{Stinson2006} and via ionizing radiation from massive stars
(early stellar feedback) before they turn into SN \citep{Stinson2013}.
Metals are produced by type II and type Ia SN.  These, along with
stellar winds from asymptotic giant branch stars also return mass to
the ISM.  The fraction of stellar mass that results in SN and winds is
determined using  the \citet{Chabrier2003} stellar initial mass
function.
We refer the reader to \citet{Wang2015} for a more
detailed description of the code and the simulations.

\vspace{-0.5cm}
\subsection{NIHAO galaxies}

Since our aim is to study the impact of baryons on the galaxy velocity
function it is very important to use realistic simulated galaxies.
In this respect, galaxies from the  NIHAO project perform extremely
well being consistent with a wide range of galaxy properties both in
the local and distant universe: evolution of stellar to halo masses
and star formation rates \citep{Wang2015}; cold and hot gas content
\citep{Stinson2015, Wang2016, Gutcke2016}; stellar disk kinematics
\citep{Obreja2016}; shallow inner dark matter density slopes of
dwarf galaxies \citep{Tollet2016}; and resolve the too-big-to-fail
problem for field galaxies \citep{Dutton2016}.
They thus provide plausible, even though not unique, templates for the
relation linking the dark matter circular velocity and the velocity
from \hi dynamics.

\vspace{-0.5cm}
\subsection{\hi calculation}

In calculating the neutral hydrogen \hi fraction we 
use the self shielding approximation described in \citet{Rahmati2013},
based on full radiative transfer simulations presented in
\citet{Pawlik2011}.  The self shielding approximation defines a
density threshold above which the photoionization of \hi is
suppressed. We calculate this threshold by interpolating between the
redshift values of our fixed UV background based on \citet{Haardt2001}
and then using Table 2 in \citet{Rahmati2013}. The overall effect of
this self-shielding approach is to increase the amount of \hi
(relative to the fiducial calculation in {\sc gasoline}) bringing the
simulations in better agreement with observations \citep{Rahmati2013, Gutcke2016}.

\begin{figure*}
\includegraphics[width=1.0\textwidth]{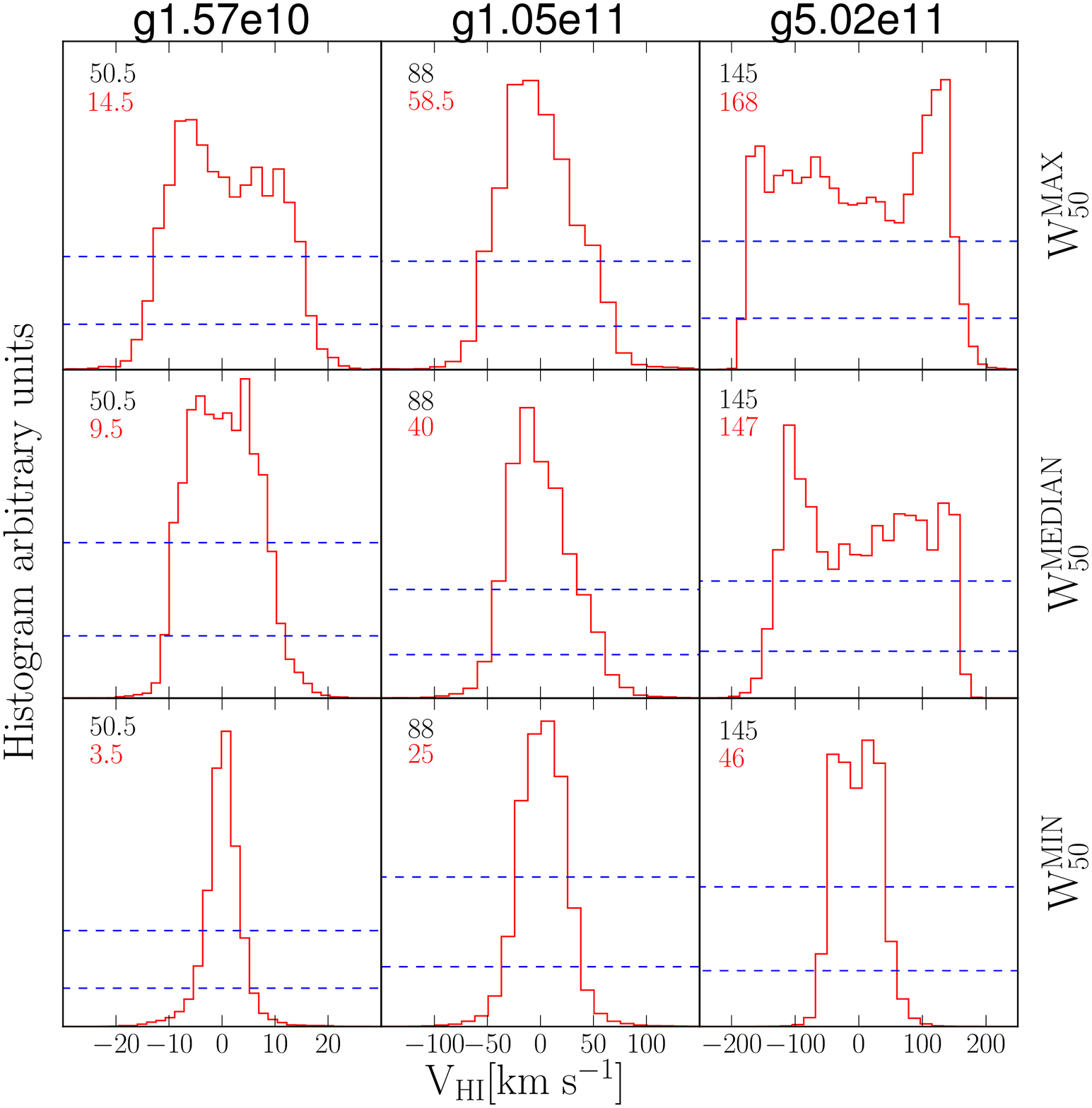}
\caption{Example integrated line-of-sight \hi velocity histograms from
  3 NIHAO galaxies from low to high mass (left to right). The upper
  panels show the projection resulting in the maximum $W_{50}$, the
  middle panels the median $W_{50}$ and the lower panels the minimum
  $W_{50}$.}
\label{fig:VHI}
\end{figure*}

\vspace{-0.5cm}
\section{The observational data}
\label{sec:obs}

In this Letter we will use the  velocity function as computed by K15
using the Updated Nearby Galaxy Catalogue from \citet{Karachentsev2013} 
which contains redshift independent distances $D<11$
Mpc for 869 galaxies. The velocities have been obtained from \hi
velocity widths for the large majority of the galaxies ($\approx
90 per cent$) while for galaxies without \hi line-width measurements they
have assigned a line of sight velocity measurement using the average
luminosity velocity relation in the K band for galaxies with measured
stellar velocity dispersions.
The resulting GVF is in reasonably good
agreement with previous pure \hi results from HIPASS \citep{Zwaan2010} and
ALFALFA \citep{Papastergis2011} in the velocity range 25-150 \kms
where they overlap.
In order to better compare with simulations K15 used
$V_{50}$ (called $V_{\rm los}$ in K15) as observational
velocity, which is defined as the width of the velocity distribution
at half height ($W_{50}$) divided by two.
We will use the original datapoints from K15 as the observational velocity function in the rest of this Letter.

\vspace{-0.5cm}
\section{Results}
\label{sec:res}
Observed rotation velocities are usually based on wide area, 21 cm
surveys. Thanks to their spectroscopic nature, \hi surveys
automatically provide the spectral \hi line profile of every detected source. 
The velocity width ($W_{50}$) of each galaxy can be easily
extracted and it is then possible to construct a velocity function
\citep[e.g.,][]{Papastergis2011}.

As a first step we have replicated this observational process
in our galaxies.  Fig.~\ref{fig:VHI} shows the velocity distribution
of the \hi gas in three different galaxies with increasing mass from
left to right. Since the \hi velocity depends on the galaxy inclination,
we show for each galaxy three different orientations that yield the
maximum  (upper row), median (middle row) or minimum (lower row)
velocity width.  The two horizontal lines indicate two commonly used
velocity definitions: the width at 50 per cent ($W_{50}$) and 20 per cent ($W_{20}$)
of the  maximum height. We calculate the maximum height as the average
between the maximum flux on the positive and negative velocity sides.
As expected, low mass galaxies tend to have a single peaked
distribution, which is roughly Gaussian,
while more massive ones show clear signs of rotational support
and have double peaked \hi profiles.

\begin{figure}
\centering
\includegraphics[width=0.45\textwidth]{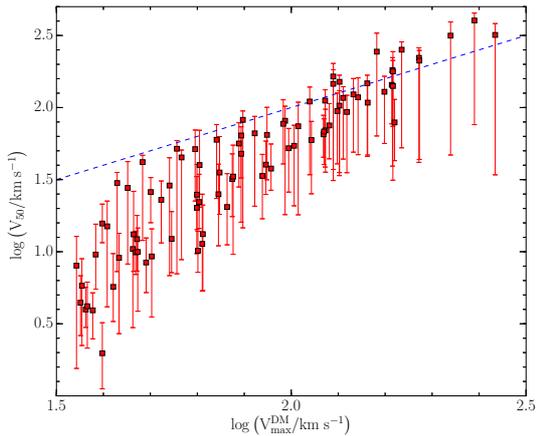}
\caption{Relation between the \hi velocity, $V_{50}\equiv W_{50}/2$,  from
  the hydro simulation and the maximum circular velocity, \Vmax, from
  the corresponding dark matter only simulation. The squares show the
  median velocity from 100 random projections, while the error bar
  connects the minimum and maximum values. The blue line corresponds to
  the 1:1 relation.}

\label{fig:V50Vmax}
\end{figure}

For each galaxy the dark matter maximum circular velocity \Vmax~
(defined as the maximum of $\sqrt{GM/R}$) of the corresponding N-body run)
is indicated in the top left corner of every panel (black), together with the
value of the rotational velocity of the galaxy $V_{50}$ (red),
defined as {\it half} of the width at half high ($V_{50} =
0.5 W_{\rm 50}$).
One can immediately notice that while for massive galaxies the two
velocities ($V_{50}$ and \Vmax) are quite similar, strong
differences are present at lower masses, where the velocity from \hi
kinematics can be even less than 20 per cent of circular velocity of the DM.
This trend with mass is clearly visible in Fig.~\ref{fig:V50Vmax},
where we plot the two velocities ($V_{50}$, \Vmax) one
against the other\footnote{The actual $V_{\rm max}$ in the hydro run might
  be different from the one from the Nbody simulation, mainly due to (baryonic)
  mass loss in low mass haloes \citep[e.g.,][]{Sawala2016}. On the other hand in this Letter
  we are interested in comparing previous results from Nbody simulations
  with actual results from hydro runs and hence we are more interested
  a comparison of \Vmax~ with $V_{50}$.}.
Each galaxy is seen from one hundred different
lines of sight, the (red) squares represent the median value among the
hundred projections and the error bar is obtained by connecting the
maximum and minimum values. Our results are very consistent with recent ones
  obtained by \citet[][see their figure 5]{Vandenbroucke2016}
despite their use of a different code and a different
feedback implementation.

\begin{figure*}
\centering
\includegraphics[width=1.0\textwidth]{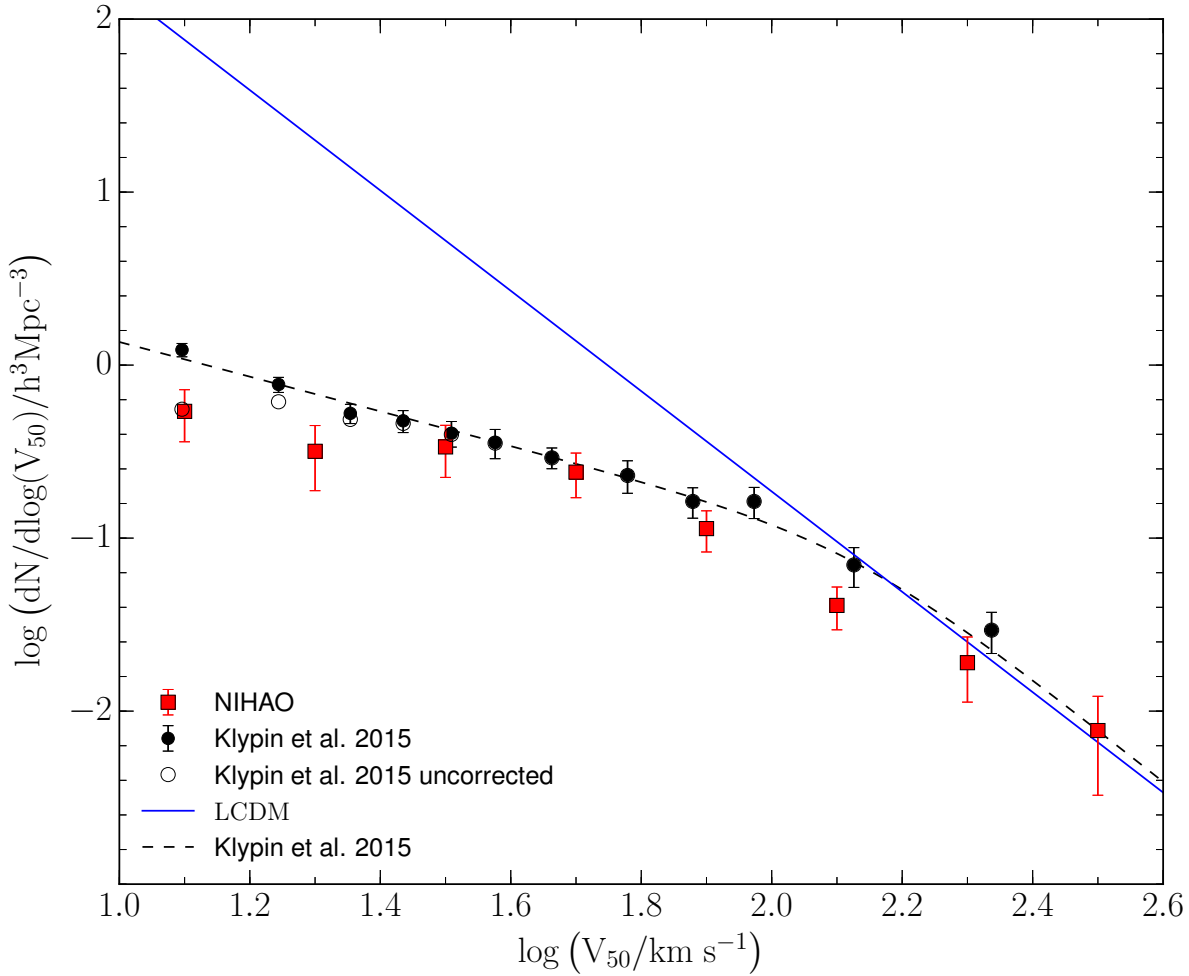}
\caption{GVF: The black points with error bars
  are the observational values from K15, the black line is an
  analytic fit to the data (K15). The blue solid line
  is the \Vmax~ prediction from pure N-body simulations from eq \ref{eq:GVF}. The red points with
  error bars are the results using \hi linewidths from the NIHAO
  simulations. } 
\label{fig:Vfunc}
\end{figure*}

\begin{figure*}
\includegraphics[width=1.0\textwidth]{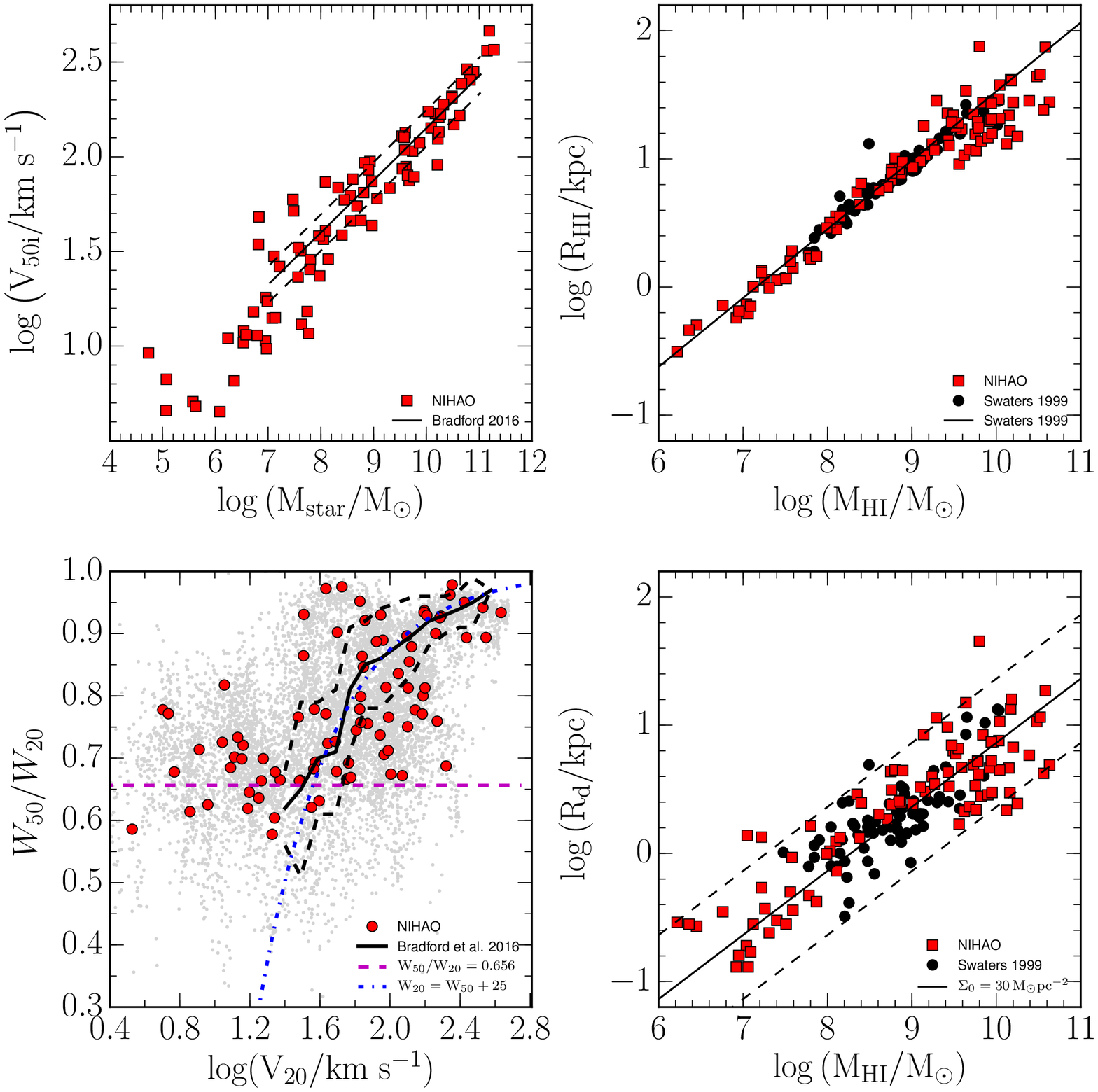}
\caption{Tully-Fisher (upper left panel), ratio of $W_{50}/W_{20}$ vs
  $V_{20}$ (lower left panel),  and \hi size vs mass (right   panels)
  relations of NIHAO simulations (red squares) compared to
  observations (black circles and curves). See text for further details.}
\label{fig:Hitest}
\end{figure*}

Large N-body simulations such as the Millennium  \citep{Springel2005} and
the Bolshoi \citep{Klypin2011} runs can be used to compute the dark
matter halo velocity distribution function, which provides the number
of haloes in a given range of \Vmax. Such a velocity function can be
well approximated by a single power law, and \citet{Klypin2011} using
the Bolshoi simulation suggested the following parameterization for a
Planck cosmology:
\begin{equation}
  { {\rm d} N \over {{\rm d} \log_{10} V_{\rm max}^{\rm DM}}} = 0.186 \left(  V_{\rm max}^{\rm DM} \over {100\, \kms}
  \right)^{-2.90} h^{3} {\rm Mpc}^{-3}.
  \label{eq:GVF}
\end{equation}

The NIHAO suite is based on high-resolution simulations of single
galaxies, for this reason we need to assign an appropriate weight for
each halo in order to compute a velocity function.  We proceeded as
follows: we first computed the number of haloes in bins of
$\log$(\Vmax) using their N-body counterpart (for a total of 87
objects).  We then associated to each \Vmax~ bin a weight, calculated
as the ratio between our results and the global ones from K15
(eq. \ref{eq:GVF}).  Finally, we apply these weights to the whole sample
of $V_{50}$ from the hydro simulations (meaning to all 100 different projections
  for each galaxy, for a total of 8700 values for $V_{50}$) resulting in a velocity
function, which can
be directly compared with the observations.  We assign an error to
each point based on the Poisson noise of that velocity bin.

Results are presented in Fig.~\ref{fig:Vfunc}, which is the key
plot of this Letter: in black we show the observed velocity function
(points) and its analytic fit via Eq.~11 in K15, in blue the dark
matter halo velocity function from the Bolshoi simulations (see
K15), and finally in red the velocity function as predicted by the
NIHAO simulations accounting for the effects of galaxy formation.
This figure clearly shows that, once the mass dependent offset between
the \hi velocity and the maximum velocity of the halo is taken into
account, hydrodynamical simulations based on the CDM
model have no problems in reproducing the observed velocity function.

The NIHAO results seem to fall a bit short at the low velocity end
when compared with the observations corrected for incompleteness
(black filled circles).  On the other hand they agree well with the
raw data (open circles).  This might indicate that the \hi gas in the
NIHAO simulations is too turbulent, or is not extended enough (but see
next section) or that the incompleteness correction applied by K15 was
a bit too generous. Deeper future surveys such as with the Square
Kilometer Array \citep{SKA} will help in clarifying this issue.

\vspace{-0.5cm}
\subsection{\hi distribution in NIHAO}
To validate our GVF results we need to show that NIHAO galaxies
have realistic \hi velocity distributions and radial extents.
To address this point we show several scaling relations.
The upper left panel of Fig.~\ref{fig:Hitest} shows 
the relation between stellar mass and \hi velocity in
NIHAO (red squares) and in observations as measured by \citet{Bradford2016}.
The velocities from \citet{Bradford2016} are corrected for
inclination.  To compute an equivalent quantity in the NIHAO galaxies
for each simulation we find the median linewidth from
the 100 projections and then divide it by $0.866=\sin(60)$, 
the median inclination correction for a random distribution of galaxies in the
sky. We have dubbed this average velocity  $V_{\rm 50i}$. The
Tully-Fisher relations from the NIHAO simulations are in very good
agreement with observation. 
%
The lower left panel of Fig.~\ref{fig:Hitest} shows the relation
between  the NIHAO linewidths measured at 50 per cent and 20 per cent of the peak
flux  (grey points for all 100 projections per galaxy and red squares
for the median velocities).  At velocities ($V_{20} > 100 \kms$) the
two linewidths are similar, reflecting the double peaked nature and
steep wings of the line profiles. However,  at lower velocities
$W_{50}/W_{20}$ decreases following $W_{20}=W_{50}+25 \,\kms$ and at
$V_{20}\simeq 30 \kms$ reaches a value consistent with that of a
Gaussian $W_{50}/W_{20}\sim 0.656$.  This change in line profile shape
contributes to $V_{50}$ deviating from \Vmax\, at low velocties
\citep{Brook2016}.  The good agreement between NIHAO simulations and
observations from 
\citet{Bradford2016} (solid black curve with 1$\sigma$ scatter) shows that NIHAO
simulations have realistic \hi line profiles.

The NIHAO galaxies are also able to reproduce the relation between the
total mass in \hi and its radial extent, as shown in the right panels
of Fig.~\ref{fig:Hitest}. In the upper one we compare the \hi mass
with the scale radius of  its radial distribution. This radius has
been obtained by fitting an exponential profile to the face-on \hi
surface density  between the radii containing 10 per cent and 90 per cent of the
\hi mass.  In the bottom right panel we use the
radius where the \hi surface  density drops below 1 $ {\rm
  M}_{\odot}$/pc$^2$ \citep[see][]{Swaters1999}.
The NIHAO galaxies
(in red) are able to reproduce the observed \hi distributions (black
points) for both definitions of the \hi radius.

\vspace{-0.5cm}
\section{Discussion and Conclusions}
\label{sec:conc}

The abundance of galaxies with a given velocity linewidth
($V_{50}\equiv W_{50}/2$) provides a very powerful tool to test
against observations different models for Dark Matter, from Cold (CDM)
to Warm (WDM).  The slope of this velocity function at low velocities
($V_{50}<100\,\kms$) departs significantly from the expectations of
the standard CDM model, leading to a difference in abundance of about
a factor $\approx 8$ at $V_{50} \sim 50$ \kms \citep[][ALFALFA
  survey]{Papastergis2011}.  Such a discrepancy is still in place when
deeper observations of the local ($D<10$ Mpc) Universe are used
\citep[K15, based on the survey from][]{Karachentsev2013}; moreover a
simple cut in the matter power spectrum, as in WDM models, does not
provide a viable solution (K15).
These kinds of analyses suffer from the fact that
\hi discs in low mass
galaxies might not be extended enough to probe the full amplitude of
the rotation curve \citep{Brook2016}.

We reinvestigated this problem using the NIHAO simulation suite to
study in detail the relation between the maximum circular velocity
of Dark Matter haloes, \Vmax,  
and the \hi velocity.
We found that at high masses the \hi rotation curves are extended
enough to fully capture the dynamical velocity of the dark matter, and
hence the linewidth based velocities, $V_{50}$, can be directly
compared with velocities coming from dissipationless (N-body)
simulations.  At lower masses, namely $\Vmax < 100\,\kms$, there is a
clear bias between the circular velocity of the halo and the width of
the \hi velocity distribution. Such a bias is due to both the \hi
distribution not being extended enough to reach the peak of the
circular velocity and to the departure of the \hi dynamics from a
purely rotating disc, due to substantial dispersion in the vertical
direction. When such a bias is taken into account, the simulated \hi
velocity function is able to match the observed ones well, showing
that the CDM model is not intrinsically flawed,
but that attempts to realistically ``observe'' fully hydrodynamical simulations
might alleviate apparent tensions between observations and
theory predictions. In conclusions any comparisons
between pure N-body simulations and observations must be taken with
a grain of salt.

\vspace{-0.5cm}
\section*{Acknowledgments}

We thank Anatoly Klypin and Jeremy Bradford for sending us their
results in electronic form.
This research was carried out on the High Performance Computing
resources at New York University Abu Dhabi; on the  {\sc theo} cluster
of the Max-Planck-Institut f\"ur Astronomie and on the {\sc hydra}
clusters at the Rechenzentrum in Garching.
%
XK is supported by the NSFC (No.11333008) and the strategic priority research program of CAS (No. XDB09000000).

\vspace{-0.5cm}
\bibliographystyle{mnras}
\bibliography{bibliography}

\end{document}